\documentclass[preprint,12pt]{elsarticle}
\usepackage{amssymb}
\usepackage{graphicx}
\usepackage{epsfig}
\journal{International Journal of Electronics and Communications}

\begin{document}
\begin{frontmatter}

\title{IMPROVED REALIZATION OF CANONICAL
CHUA'S CIRCUIT WITH SYNTHETIC INDUCTOR USING CURRENT FEEDBACK
OPERATIONAL AMPLIFIERS}

\author{ R. JOTHIMURUGAN$^\dag$, K. SURESH$^\S$,\\ P. MEGAVARNA EZHILARASU$^\ddag$ and K. THAMILMARAN$^{*}$}

\address{Centre for Nonlinear Dynamics, School of Physics, \\
Bharathidasan University, Tiruchirappalli - 620 024, India\\
$^\dag$jothi.nld@gmail.com\\$^\S$sureshscience@gmail.com
\\$^\ddag$megaezhil77@gmail.com\\$^{*}$maran.cnld@gmail.com}

\begin{abstract}
In this paper, we report an improved implementation of an inductorless third order autonomous canonical Chua's circuit. The active elements as well as the synthetic inductor employed in this circuit are designed using current feedback operational amplifiers (CFOAs). The reason for employing CFOAs is that they have better features such as high slew rate, high speed of operation, etc., which enable the circuit to operate at higher frequency ranges, when compared to the circuits designed using voltage operational amplifiers. In addition to this, the inclusion of CFOAs provide a buffered output which directly represent a state variable of the system. The Multisim simulations in the time and frequency domains confirm the theoretical estimates of the performance of the proposed circuit at high frequencies. It is also confirmed through hardware experiments. 
\end{abstract}

\end{frontmatter}

\section{Introduction}	

Since the later part of nineteenth century, the field of chaos has attracted intense interest and active investigations. This may be attributed to the exciting applications that chaos finds in such diverse fields as engineering, communications, bio-medicine, chemical kinetics,  neural networks, visual sensing, music, etc \cite{1,2,28}. Research in this emerging field has been aided to a great extent by nonlinear electronic circuits, especially those having piecewise-linear nonlinearities. Prominent among the piecewise-linear circuits 
are the Chua's circuit family employing the well known Chua's diode as their nonlinear element \cite{3,4}. The reason for the popularity of these circuits is that their nonlinear element, namely Chua's diode can be realized easily using simple circuit elements while at the same time it is also mathematically tractable \cite{5}. Different circuits in this Chua's family exhibit different dynamical behaviours based on parameters as well as the configuration of the Chua's diode in the circuit. However even among the Chua's circuit family, the canonical Chua's circuit occupies a prime position. This is because the canonical Chua's circuit can generate all possible phenomena associated with any three region symmetric piecewise-linear continuous vector field while at the same time it contains a minimum ({\it six}) number of circuit elements \cite{7,8,9}. Yet in spite of their versatility, the applications of Chua's circuit family are limited to low frequency ranges. Also the presence of the discrete inductor makes them impossible for fabrication on ICs. To overcome these difficulties as well as to improve the frequency performance, two different approaches have been followed. First one is to redesign Chua's diode using current feedback operational amplifiers (CFOAs) \cite{10,11}. This is based on the fact that CFOAs have high slew rate and high speed operations \cite{12}. High frequency operation of the Chua's circuit, necessitates scaling down of inductance and capacitance values to impractically low levels. While very low levels of capacitor values can be realized, those of inductances are not possible. So the discrete inductor is to be replaced with synthetic inductance simulator. Hence the second approach is to construct Chua's circuit with both synthetic inductor and the Chua's diode designed using CFOAs \cite{13,16}. While these modifications for high frequency operation have been implemented for the Chua's oscillator, to the best of our knowledge no such high frequency modifications for canonical Chua's circuit is available till date. Hence in this paper, we propose an improved realization of canonical Chua's circuit with active elements and synthetic inductor using CFOAs. 

This paper is organized as follows. In section 2, the classical Chua's circuit and its CFOA based implementations, are discussed. The voltage operational amplifier based canonical Chua's circuit is described briefly in section 3 along with the simulation results. The implementation of the improved canonical Chua's circuit is described in section 4. The realization of active elements used in this circuit are also explained in detail here. A theoretical estimation of the maximum frequency of operation of this circuit is made in section 5. The results of Multisim simulations, confirming the theoretical estimates of high frequency operations, are presented in section 6. Experimental realization of the circuit and the performance of the circuit at high frequencies are discussed in section 7. Finally in
section 8, the importance of this circuit and scope for further developments are discussed.
 
\section{Classical Chua's circuit and its CFOA Realization}

In this section and the next, we describe the realizations of the classical Chua's circuit, the CFOA based Chua's circuit and the canonical Chua's circuit with the view to understand the differences in their topologies. The classical Chua's circuit shown in Fig. 1(a), is a  three dimensional piecewise-linear autonomous circuit consisting of a linear resistor, three linear dynamic elements (two capacitors and an inductor) and the Chua's diode as its nonlinear resistor ($N_{R}$). The Chua's diode is realized experimentally using two operational amplifiers and six linear resistors as shown in Fig. 1(b). It is characterized by a five-segment odd symmetric piecewise-linear voltage-current ({\large \it v$-$i}) characteristic (Fig. 1(c)). From Fig. 1(b), we find that this Chua's diode is a configuration of two negative impedance converters connected in parallel. For the standard values of resistors shown in Fig.1(b), the slopes of the characteristic curve (Fig. 1(c)) are $G_{a}=-0.756$ mA/V, $G_{b}=-0.409$ mA/V and $G_{c}=4.6~\mu$A/V. The breakpoints are $B_{p_{1}}\approx\pm 1.08$ V and $B_{p_{2}}\approx\pm 7.61$ V \cite{2,5}. However, it is to be noted that even though the Chua's diode has a five segment characteristic, the circuit operation lies in the negative slope regions only \cite{6}. The dynamics of this classical Chua's circuit have been studied
extensively using numerical simulations and hardware experiments \cite{1,4}. The same classical Chua's circuit has been modified by replacing the Chua's diode and discrete inductor coil with CFOA based nonlinearity and synthetic inductor. The dynamics of this improved
Chua's circuit has been extensively studied experimentally in Ref. \cite{10,11,13,16}.

\section{Canonical Chua's circuit}	

The canonical Chua's circuit realized using voltage operational amplifiers (VOAs) is shown in Fig. 2(a). This is also a three dimensional 
piecewise-linear autonomous circuit, consisting of two capacitors and an inductor. It has two active elements namely a nonlinear resistor ($R_{N}$) and a negative conductor ($g_{N}$). The nonlinear resistor ($R_{N}$), shown in Fig. 2b(i)  is conceived by connecting a negative impedance converter in parallel with a linear resistor resulting in a voltage-current characteristic curve having three distinct regions separated by two symmetrically placed breakpoints ($\pm B_p$). The central region between the two breakpoints has a negative slope while the other two outer regions have positive slopes. These are shown in Fig. 2b(ii). For a given set of resistor values, ($R_{1}=4.1$ k$\Omega$,
$R_{2}=19.0$ k$\Omega$, $R_{3}=1.0$ k$\Omega$ and $R_{4}=220~\Omega$) the slopes of the characteristic curve are $G_{a}=-0.8818$ mA/V, $G_{b}=4.37$ mA/V and break points are $B_{p}\approx\pm 0.581$ V. The negative conductor is just a negative impedance converter designed using operational amplifier and three resistors as shown in Fig. 2c(i). Its characteristic curve is  shown in Fig. 2c(ii). The slope of the negative conductance region is $g_{N}=-0.5$ mA/V for the resistor values $R_{1}=R_{2}=R_{3}=2.0$ k$\Omega$. 

By choosing the parameters of the canonical Chua's circuit as $C_{1}=33$ nF, $L=120$ mH, $R=355{~\Omega}$ and varying the $C_{2}$ from 100 nF to 40 nF, the circuit exhibits the standard period doubling route to chaos. Intermittency and other dynamical phenomena are also observed for a different set of parametric values \cite{8,9}. Keeping the parameters fixed as mentioned above and having $C_{2}=60$ nF, a double band chaotic attractor shown in Fig. 3(a) is observed. The power spectrum of the variable $v_{1}$ for the double band chaotic attractor is shown in Fig. 3(b). From this power spectrum, the operating frequency of the circuit is estimated to be 2.69 kHz. This low range of frequency of operation is further confirmed by implementing VOA based canonical Chua's circuit using State Controlled Cellular Neural Network (SC-CNN). In this study, the operation frequency of the circuit was centered around 2 kHz for single band chaos and 0.9 kHz for double band chaos \cite{27}.

\section{Improved canonical Chua's circuit}
Though the canonical Chua's circuit is truly canonical, in that it exhibits all the behaviours that are possible with any three region symmetric piecewise-linear continuous vector field, its main limitation is that of low maximal frequency of operation. This is attributed to the limitations of VOAs such as low slew rate, fixed gain-bandwidth product, etc \cite{28}. An improved canonical Chua's circuit that overcomes these limitations can be had if  two modifications are made. Firstly, the two active elements, namely negative conductor ($g_{N}$) and nonlinear resistor ($N_{R}$) are realized using four terminal current feedback operational amplifiers (CFOAs). Secondly, the inductance coil is replaced with a synthetic inductor ($L_{eq}$) realized using CFOAs. The CFOAs are chosen because of their high speed and high slew rate which resulting in high frequency of operation \cite{12}. This improved canonical Chua's circuit so constructed is given in Fig. 4.

In the following subsections, we describe the realization of the CFOA based active elements ($g_{N},~N_{R}$) as well as synthetic inductor $(L_{eq})$ used in the circuit. During these circuit implementations, AD844AN type CFOAs biased with $12$ V DC dual power supply were used.

\subsection{Negative Conductor ($g_{N}$)}
The negative conductance is obtained from the linear negative region of the negative impedance converter.~Since, canonical Chua's circuit operates only in the linear region of the NIC, the CFOA used here are a cascade of second generation current conveyors (CCII). This CCIIs are generally realized by effecting a direct current feedback to the non-inverting terminal of an operational amplifier and connecting a grounded resistor from its inverting terminal, as shown in Fig. 5a(i) \cite{17}. In this negative conductor sub-circuit, output voltage terminal ($v_{0}$) is kept open. This results in a direct flow of voltage across the non-inverting terminal. The characteristic curve that results in the  ({\large \it v$-$i}) plane is given in Fig. 5a(ii). The negative slope of this sub-circuit can be depicted from the expression
\begin{eqnarray}
g_{N} &=& -\frac{1}{R_{5}}. \label{eq1}
\end{eqnarray}
For the value of the resistance ($R_{5} = 2.0$ k$\Omega$), the value of slope, obtained through simulation experiment is $g_{N}=-0.5$ mA/V. This agrees well with the theoretical prediction. This CFOA based design of negative conductor circuit results in a reduction in the number of passive resistors by two, when compared to the original VOA based realizations which were used earlier.
 
\subsection{Nonlinear Resistor ($R_{N}$)}
The nonlinear resistor ($R_{N}$) in Fig. 2b(i) is designed using a conventional operational amplifier and four resistors. It operates as a voltage  controlled negative impedance converter. This sub-circuit is reconstructed using current feedback operational amplifier. In this realization shown in Fig. 5b(i), the current terminal C is kept open or equivalently output current is equal to zero. This causes  the current feedback operational amplifier to work as a voltage operational amplifier, whose characteristic has one negative and two positive slopes and two breakpoints as shown in Fig 5b(ii). The expressions for the slopes and break points are given by \cite{18},
\begin{eqnarray}
G_{a} &=& \frac{1}{R_{4}} -\frac{R_{2}}{R_{1}R_{3}} \label{eq2}\\
G_{b} &=& \frac{1}{R_{1}} +\frac{1}{R_{4}} \label{eq3}\\
\pm B_{p}&=&\frac{R_{3}}{R_{2}+R_{3}}(\pm E_{sat}).\label{eq4}
\end{eqnarray}

By choosing the values of  the various resistors as,  $R_{1}=3.3$ k$\Omega$, $R_{2}=19$ k$\Omega$, $R_{3}=1.5$ k$\Omega$ and $R_{4}=400~\Omega$ and using Eqns. (2 - 4), the slopes and break points are given as $G_{a}=-1.338$ mA/V, $G_{b}=2.803$ mA/V and $B_{p}=\pm~0.538$ V. The Multisim simulation of the circuit with the above mentioned parametric values also results in an identical characteristic curve with slopes and break points given by  $G_{a}=-1.35$ mA/V, $G_{b}=2.63$ mA/V and $B_{p}=\pm~0.588$ V. As is to be expected, there is close matching between theoretical and circuit simulation results.  

\subsection{Floating Inductance Simulator ($L_{eq}$)}

Though the inductor is an important passive device, it is a less desirable element \cite{19}, because of its inability to lend itself to integration. Hence the spiral inductor is widely used during the implementation of the integrated circuits. Even this has some drawbacks such as low tunability, large space, weight and  high cost \cite{20}. Hence to overcome these difficulties, inductance simulators are used as substitutes for inductor coils in many circuit applications. There are two major kinds of inductance simulators, namely ({\it i}) the grounded inductance simulator wherein one of its terminals is grounded and ({\it ii}) the floating inductance simulator wherein neither of the terminals are grounded. Both of these are generally used in the realization of electrical circuits. A large number of circuit realizations for floating inductance simulators have been reported in the literature \cite{21,22,23,24,25}. In this present study, the CFOA based floating inductance simulator developed by Senani has been used \cite{25}. This is because the CFOAs are proven to be quite useful in either current or voltage mode signal processing circuits \cite{20}. The inductance simulator so obtained is shown in Fig. 5(c). It is a combination of three CFOAs with two passive resistors and a capacitor connected in both voltage and current modes. It yields the equivalent inductance as
\begin{eqnarray}
L_{eq}&=&R_{1}R_{2}C.
\label{eq:ind}
\end{eqnarray}
The following elements value are chosen as $R_{1}=R_{2}=1.0$ k$\Omega$ and $C=180$ nF/$180$ pF to simulate $L_{eq}=180$ mH/$180~\mu$H. The incorporation of these CFOA based elements in the improved canonical Chua's circuit has enhanced its frequency of operation. This has been estimated theoretically in next section as well as proved from Multisim simulation in section 6.

\section{Circuit Equations: Estimation of Frequency of operation} 
As we are focusing on the enhancement of operating frequency of the canonical Chua's circuit, it is desirable that we estimate the frequency of  operation theoretically.  This can be done by writing down the dynamical equations of the canonical Chua's circuit (Fig. 4) and obtaining its characteristic equation. The circuit equations can be obtained by applying Kirchoff's laws \cite{7}. They are

\begin{eqnarray}
C_{1}\frac{dv_{1}}{dt}&=&i_{L}-g(v_{1}) \label{canoeq1}\\
C_{2}\frac{dv_{2}}{dt}&=&-g_{N}v_{2}+i_{L} \label{canoeq2}\\
L\frac{di_{L}}{dt}&=&-(v_{2}+v_{1}+i_{L}R), \label{canoeq3}
\end{eqnarray}

{\noindent where $L$ represents the simulated inductance ($L_{eq}$) and $g(v_{1})$ represents the characteristic of the nonlinear resistor ($R_{N}$) whose mathematical form is,}
\begin{eqnarray}
g(v_{1})&=&G_{b}v_{1}+0.5(G_{a}-G_{b})[|v_{1}+B_{p}|-|v_{1}-B_{p}|].
\label{eq:cano1}
\end{eqnarray}

{\noindent Here, $G_{a}$ and $G_{b}$ are the inner negative and outer positive slopes of the three segment piecewise-linear characteristic of nonlinear resistor and $\pm B_{p}$ represents the break points.}

The characteristic equation of the linearized canonical Chua's circuit corresponding to the inner region can be written by
\begin{equation}
s^{3}+a_{2}s^{2}+a_{1}s+a_{0}=0
\label{eq:li_cano}
\end{equation} 
where 
\begin{eqnarray}
a_{2}&=& \frac{g_{N}}{C_{2}}+\frac{R}{L}+\frac{G_{a}}{C_{1}}\\
a_{1}&=&\frac{g_{N}R}{LC_{2}}+\frac{G_{a}g_{N}}{C_{1}C_{2}}+\frac{G_{a}R}{LC_{1}}+\frac{1}{LC_{1}}+\frac{1}{LC_{2}}\\
a_{0}&=&\frac{G_{a}g_{N}R}{LC_{1}C_{2}}+\frac{G_{a}}{LC_{1}C_{2}}+\frac{g_{N}}{LC_{1}C_{2}}.
\label{eq:coeff}
\end{eqnarray}
Similarly, the characteristic equation for the two outer segments can be obtained as,
\begin{equation}
s^{3}+a^{'}_{2}s^{2}+a^{'}_{1}s+a^{'}_{0}=0
\end{equation} 
 The corresponding coefficients in these cases are given as,
\begin{eqnarray}
a^{'}_{2}&=& \frac{g_{N}}{C_{2}}+\frac{R}{L}+\frac{G_{b}}{C_{1}}\\
a^{'}_{1}&=&\frac{g_{N}R}{LC_{2}}+\frac{G_{b}g_{N}}{C_{1}C_{2}}+\frac{G_{b}R}{LC_{1}}+\frac{1}{LC_{1}}+\frac{1}{LC_{2}}\\
a^{'}_{0}&=&\frac{G_{b}g_{N}R}{LC_{1}C_{2}}+\frac{G_{b}}{LC_{1}C_{2}}+\frac{g_{N}}{LC_{1}C_{2}}.
\label{eq:coeff1}
\end{eqnarray}
The condition for oscillation is $a_{1}a_{2}=a_{0}$ for the central region  and correspondingly $a^{'}_{1}a^{'}_{2}=a^{'}_{0}$ for the outer regions. The radian frequency of oscillation is found to vary in the range ($\sqrt{a^{'}_{1}} \le \omega_{0} <\sqrt{a_{1}}$).
Or alternatively, the linear frequency of oscillation is given as
\begin{equation}
\frac{1}{2\pi}\sqrt{a^{'}_{1}} \le f_{0} < \frac{1}{2\pi}\sqrt{a_{1}}.
\label{eq:real_fre}
\end{equation}

For the specific choice of parameters, $C_{1}=7.5$ nF, $C_{2}=60$ nF, $L_{eq}=180$ mH, $R=700~\Omega$, $g_{N}=-0.5$ mA/V and $G_{a}=-1.35$ mA/V, for the inner region, the frequency of oscillation calculated using the Eqn. (\ref{eq:real_fre}) is $6.0479$ MHz. However for the outer region, as the slope takes on a positive value, namely $G_{b}=2.63$ mA/V, no oscillations will be generated \cite{26}. Only an exponential decay of any oscillatory energy is seen. This means that the frequency of oscillations will be $0$ MHz in the outer regions. Hence, the central operating frequency of the circuit is expected to be within these lower and upper bounds, that is, $0$ MHz $\le f_0 < 6.0479$ MHz.

\section{Simulation Results}
To validate our theoretical predictions, we have also studied the circuit by simulating its behaviour using Multisim - a popular Spice based circuit simulation software. The Multisim model of the proposed circuit is shown in Fig. 4. Here $g_{N},~R_{N}$ and $L_{eq}$ refer to the negative conductor, nonlinear resistor and synthetic inductor respectively.  The voltage output terminal in the negative conductor is kept free.  It is the direct buffered isolated output of the state variable $v_{2}$ of the circuit connected to the non-inverting input of the negative conductor $g_{N}$. The circuit parameters $C_{1}$, $C_{2}$ and $L_{eq}$ are fixed as constant values and the series linear resistor $R$ is used as control parameter. In order to observe the performance of the circuit in low as well as high frequency regions, we have chosen two sets of parameters. 

For the low frequency range, the parameter values are fixed as $C_{1}=7.5$ nF, $C_{2}=60$ nF and $L_{eq}=180$ mH as is the case for a conventional VOA based circuit. By decreasing the control parameter value $R$ from $1000~\Omega$, we observed the standard period doubling bifurcation route to chaos. The double band chaotic dynamics of the circuit obtained when $R=900~\Omega$ was found to have an operating frequency centered around 760 Hz.
For the high frequency range, the parameters of the circuit was scaled down by the factor 1000. This resulted in the reduction of the parameter values. The reduced values are $C_{1}=7.5$ pF, $C_{2}=60$ pF and $L_{eq}=180~\mu$H.

Similar to the low frequency operation, here also when the control parameter  $R$ was decreased in value from 1000$~\Omega$, the circuit was found to exhibit the standard period doubling route to chaos. Fig. 6 shows a sequence of phase portraits in the ($v_{1}-v_{2}$) plane depicting the period doubling towards chaos. The power spectrum for the circuit variables $v_{1}$ and $v_{2}$ for single band and double band chaotic dynamics are given in Fig. 7. From these power spectra, it is clearly seen that the dominant operating frequency of the system is centered around 1.0313 MHz for single band chaos and around 1.000 MHz for double band chaos.These dominant frequencies ($\approx 1$ MHz) obtained from Multisim simulations, are well within the expected frequency range that has been predicted theoretically. 

To show the dynamical transitions of the improved circuit, we plot the one parameter bifurcation diagram in the ($v_1-R$) plane using Multisim simulation. The circuit parameters are fixed as $C_1=7.5$ pF, $C_2=60$ pF, $L_{eq}=180~\mu$H, $G_N=-0.5$ mS, $G_a=-1.338$ mA/V, $G_b=2.803$ mA/V, $B_p=\pm 0.588$ V and set the initial values for the capacitors ($C_1, C_2, C$) with voltages ($4$ V, $3$ V, $2.5$ V). The time series at voltage $v_1$ with time step $\Delta t=10$ nS is stored for the range between $0.25$ mS to $0.35$ mS by decreasing the control parameter $R$ from $1000~\Omega$ to $800~\Omega$ with the step of $1~\Omega$. This stored data is then processed numerically and the result is plotted in Fig. \ref{fig_bif}. The figure shows the familiar period-doubling bifurcation sequence to chaos followed by periodic windows, intermittency and so on. This confirms the behaviour of the circuit which is unaltered while increasing the operating frequency of the circuit.

\section{Experimental Results}
The hardware implementation of the improved canonical Chua's circuit is made and realized in PCB is shown in Fig. \ref{fig9}. The PCB layout and the PCB circuit itself are shown in Fig.\ref{fig9}(a) and (b) respectively. This design involves five operational amplifiers, seven resistors, three capacitors and a potentiometer as a variable resistance. This variable resistance of the potentiometer is taken as control parameter. This circuit is biased with $\pm12~$V dual power supply. The op-amp used in this design are AD844AN type. We consider the voltages $v_1$ and $v_2$ to observe the performance of the circuit on the CRO. 

To show the operation of the circuit (shown in Fig. \ref{fig9}), in the high frequency range, we choose the ``off-the-shelf" components. The values of the components are $C_1=68~$pF, $C_2=6.8~$pF, $C_3=180~$pF, $R_1=1.990~{\mathrm{k}}\Omega$, $R_2=983~\Omega$, $R_3=976~\Omega$, $R_4=3.261~{\mathrm{k}}\Omega$, $R_5=386~\Omega$, $R_6=1.479~{\mathrm{k}}\Omega$ and $R_7=18.06~{\mathrm{k}}\Omega$. These alues have been measured using APLAB MT4080A hand held LCR meter. For the parameter values chosen, the value of the simulated inductance tuns out to be $L_{eq}=R_2R_3C_3=172.7\mu$H. When the resistance of the potentiometer $R$ is varied from higher to lower values, the circuit exhibits the familiar period doubling route to chaos. The experimental phase portraits depicting these are captured using analog oscilloscopes while the time series and the power spectrum analysis are made using AGILENT MSO 6014A. The experimental results are displayed in Fig. \ref{fig10}. The phase portraits in the ($v_1-v_2$) plane for periodic (period -1 limit cycle) and chaotic (single band chaos) motions are shown in panels $a(i)$ and $b(i)$ for the resistance $R$ values $471~\Omega$ and $363~\Omega$ respectively. The time series of the voltages $v_1$ (yellow line) and $v_2$ (green line) sampled at the rate of 2GSa/S are given in panels $a(ii)$ and $b(ii)$ for periodic and chaotic motion. The power spectrum of the variable $v_1$ is taken with the sampling rate of 5MSa/S to estimate the frequency of operation of the circuit. The $v_1$ has the dominant peak at 706 kHz for periodic and 810 kHz for chaotic motion respectively. These are depicted in the panels $a(iii)$ and $b(iii)$.
    
\section{Conclusion}
An improved implementation of canonical Chua's circuit with CFOA based active elements ($g_{N},~N_{R}$) and synthetic inductor ($L_{eq}$) is reported.
This improved implementation is said to have many advantages when compared to the classical design of the canonical circuit. Firstly, the inclusion of CFOA based active elements and scaling down of the circuit parameters have resulted in the enhancement of range of operating frequency of the circuit. This has been predicted theoretically, proved through simulations and confirmed by hardware experiments. Secondly, the replacement of the inductor coil with synthetic inductor makes the circuit amenable for fabrication of integrated circuits (ICs) and hence in the study of coupled dynamics and spatio-temporal chaos. Thirdly, the improved realization results in a reduction of the number of circuit components by two. 
Fourthly, the CFOA based realization of negative conductor makes available directly one of the circuit variables as a buffered output. This fact can be used effectively  in the design of chaotic cryptography systems or in any circuit wherein a chaotic output is needed. However, this circuit has a limitation too. The value of the capacitances must never be reduced below 5.0 pF so as to avoid parasitic capacitative oscillations. 

In addition to these, the improved canonical Chua's circuit exhibits, apart from the period doubling bifurcation sequence, other dynamical behaviours. For example, the intermittency route to chaos has been observed in this circuit for a different range of the control parameter. The authors intend to explore this and other behaviours in this circuit. Much progress has already been made and the results will be published later.
 
\section{Acknowledgments}
KT acknowledges the Department of Science and Technology (DST) of India, for the financial support through the grant no. SR/S2/HEP-015/2010 and RJ thanks the University Grants Commission of India for the research fellowship through the UGC-RFSMS grant no. 29427/E12/2008.

\newpage
\begin{center}
	FIGURE CAPTIONS
\end{center}	
\begin{enumerate}	
	\item The schematic representation of (a) classical Chua's circuit, $N_{R}$ represents Chua's diode, (b) realization of the VOA based Chua's diode sub-circuit and its $(v-i)$ characteristic is shown in (c).
	
	\vspace*{23pt}
	\item The schematic representation of (a) canonical Chua's circuit and  voltage operational amplifier based (i) circuit realization and (ii) ({\it v$-$i}) characteristics  of (b) Nonlinear resistor ($R_{N}$) and (c) Negative conductor ($g_{N}$).
	
	\vspace*{23pt}
	\item Double band chaotic dynamics of conventional VOA based canonical Chua's circuit. (a) Phase portrait in the ($v_{1}-v_{2}$) plane and (b) Power Spectrum for the time series of the circuit variable $v_{2}$, where the operating frequency is centered around 2.69 kHz.
	
	\vspace*{23pt}
	\item The improved implementation of canonical Chua's circuit with active elements $g_{N}$, $R_{N}$ and synthetic inductance simulator $L_{eq}$. The sub-circuit to realize $L_{eq}$ is given in Fig. 5(c). 
	
	\vspace*{23pt}
	\item (i) Current feedback operational amplifier (CFOA) based circuit realization and (ii) the simulated ($v-i$) characteristics of (a) Negative conductor ($g_{N}$), (b) Nonlinear resistor ($R_{N}$) and (c) the synthetic floating inductance simulator ($L_{eq}$) realized using CFOAs.
	
	\vspace*{23pt}
	\item The period doubling scenario in the improved canonical Chua's circuit shown in Fig. (4). The phase portraits in the ($v_{1}-v_{2}$) plane, for the fixed values of the circuit elements chosen as $C_{1}=7.5$ pF, $C_{2}=60$ pF and $L_{eq}=180~\mu$H, by decreasing resistance ($R$) values: (a) period - 1 limit cycle ($R=985~\Omega$), (b) period - 2 limit cycle ($R=968~\Omega$), (c) period - 4 limit cycle ($R=960~\Omega$), (d) single band chaotic attractor ($R=900~\Omega$), (e) double band chaotic attractor ($R=850~\Omega$) and (f) boundary ($R=805~\Omega$).
	
	\vspace*{23pt}
	\item Power spectrum for the time series of the variables $v_{1}$ and $v_{2}$ of the canonical Chua's circuit. The dominant frequency is centered around (a-b) 1.0313 MHz for the single band chaos and (c-d) 1.000 MHz for the double band chaos.	

	\vspace*{23pt}
	\item The one parameter bifurcation diagram generated in the ($v_1-R$) plane for the time series of the variable $v_{1}$. The filled circles in the horizontal axis denotes the parameters used to plot the phase portraits in Fig. \ref{fig_phase}.
	
	\vspace*{23pt}
	\item (color online) The canonical Chua's circuit (a) PCB layout and (b) the PCB circuit itself.

	\vspace*{23pt}
	\item (color online) The phase portrait in ($v_1-v_2$) plane for $a(i)$ periodic $b(i)$ chaotic motion for the value of the resistor $R$ as $471~\Omega$ and $363~\Omega$ respectively. The time series for the variable $v_1$ (yellow) and $v_2$ (green) are shown in $a(ii)$ for periodic and $b(ii)$ d for chaotic motion. Scale:- $x-$axis: $2\mu$S/Div; $y-$axis: (a) voltage $v_1$-$1$V/Div and (b) voltage $v_2$-$5$V/Div. The power spectrum of $v_1$ showing the dominant peak at 706kHz for periodic and 810 kHz for chaotic motion are depicted in $a(iii)$ and $b(iii)$ respectively. The amplitude of these two dominant peaks are noted as 5.7 dBV and 5.2 dBV.

\end{enumerate}

\newpage
\begin{figure}[th]
\centerline{\psfig{file=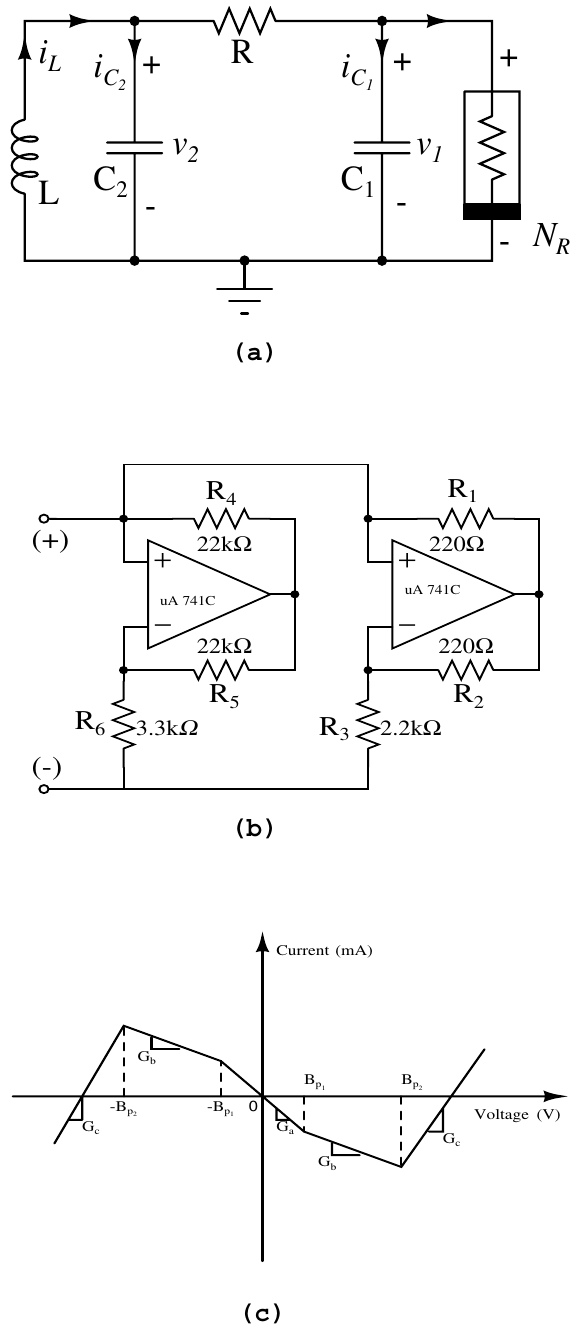,width=3.0in}}
\vspace*{6pt}
\caption{The schematic representation of (a) classical Chua's circuit, $N_{R}$ represents Chua's diode, (b) realization of the VOA based Chua's diode sub-circuit and its $(v-i)$ characteristic is shown in (c).}
\end{figure}

\newpage
\begin{figure}[th]
\centerline{\psfig{file=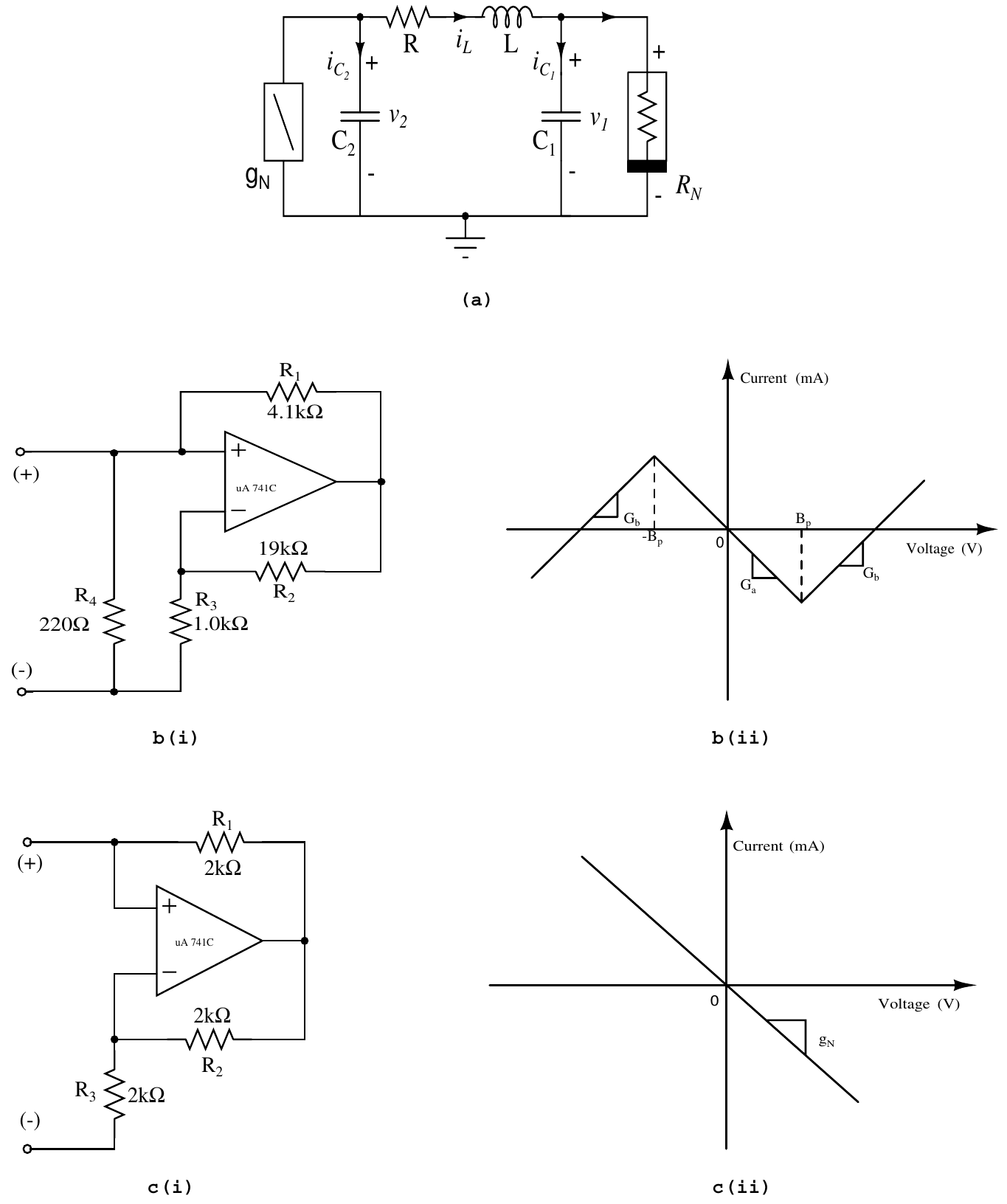,width=5.5in}}
\vspace*{6pt}
\caption{The schematic representation of (a) canonical Chua's circuit and  voltage operational amplifier based (i) circuit realization and (ii) ({\it v$-$i}) characteristics  of (b) Nonlinear resistor ($R_{N}$) and (c) Negative conductor ($g_{N}$).}
\end{figure}

\newpage
\begin{figure}[th]
\centerline{\psfig{file=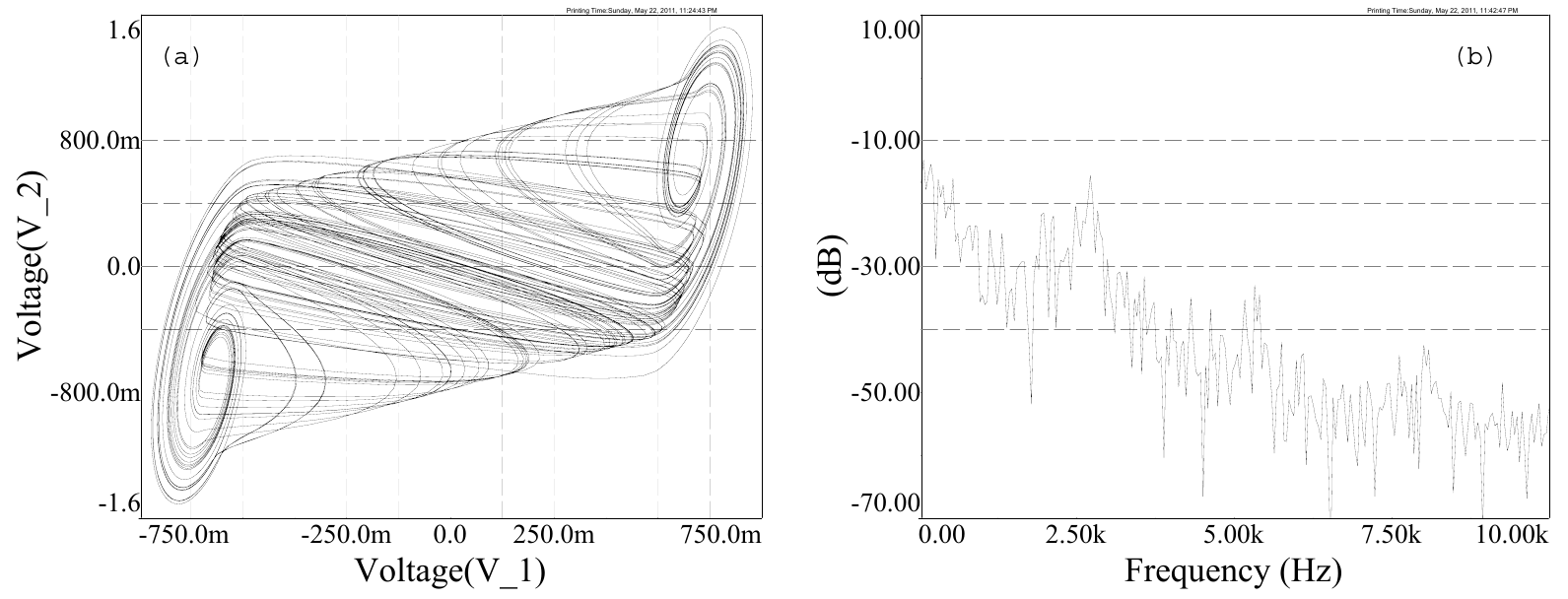,width=6.0in}}
\vspace*{6pt}
\caption{Double band chaotic dynamics of conventional VOA based canonical Chua's circuit. (a) Phase portrait in the ($v_{1}-v_{2}$) plane and (b) Power Spectrum for the time series of the circuit variable $v_{2}$, where the operating frequency is centered around 2.69 kHz.}
\end{figure}

\newpage
\begin{figure}[th]
\centerline{\psfig{file=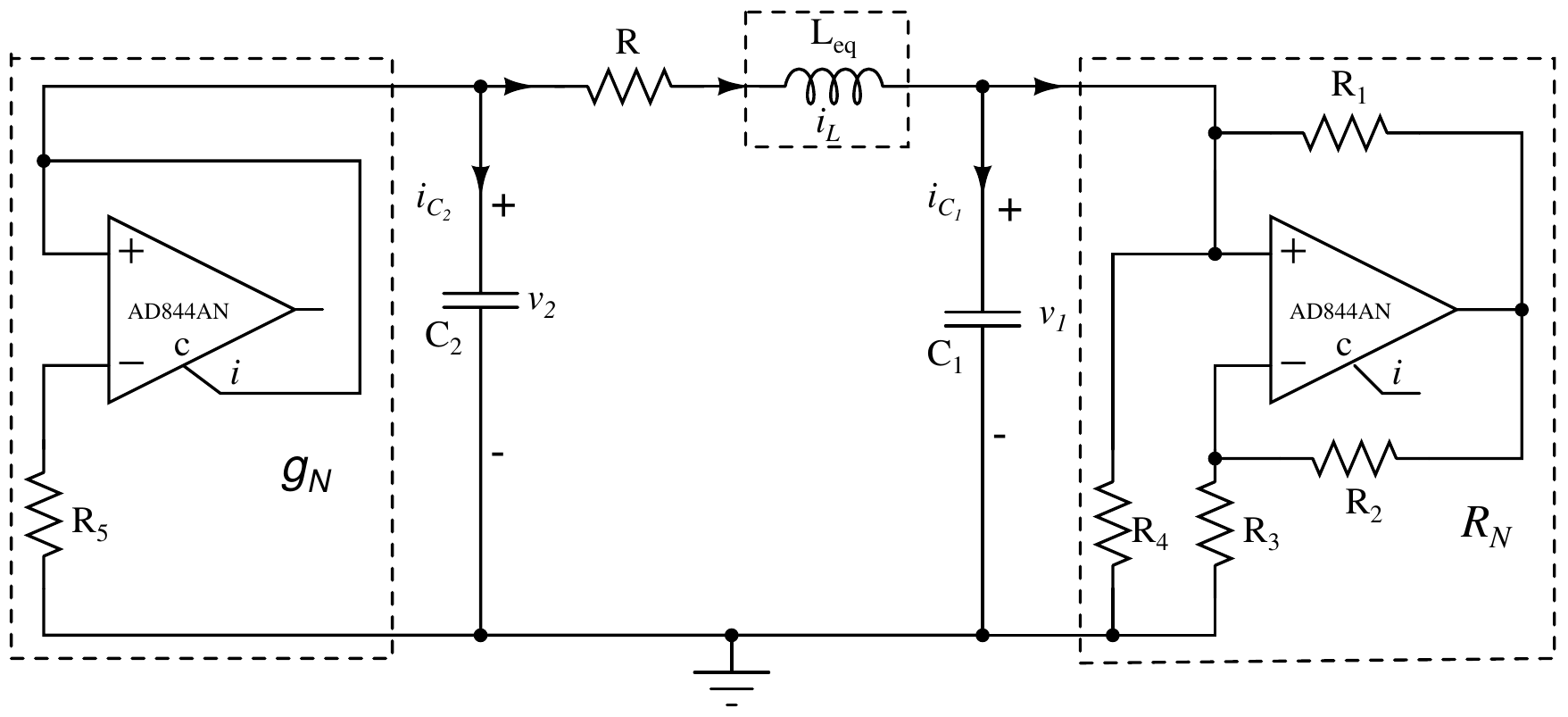,width=6.0in}}
\vspace*{6pt}
\caption{The improved implementation of canonical Chua's circuit with active elements $g_{N}$, $R_{N}$ and synthetic inductance simulator $L_{eq}$. The sub-circuit to realize $L_{eq}$ is given in Fig. 5(c).}
\end{figure}

\newpage
\begin{figure}[th]
\centerline{\psfig{file=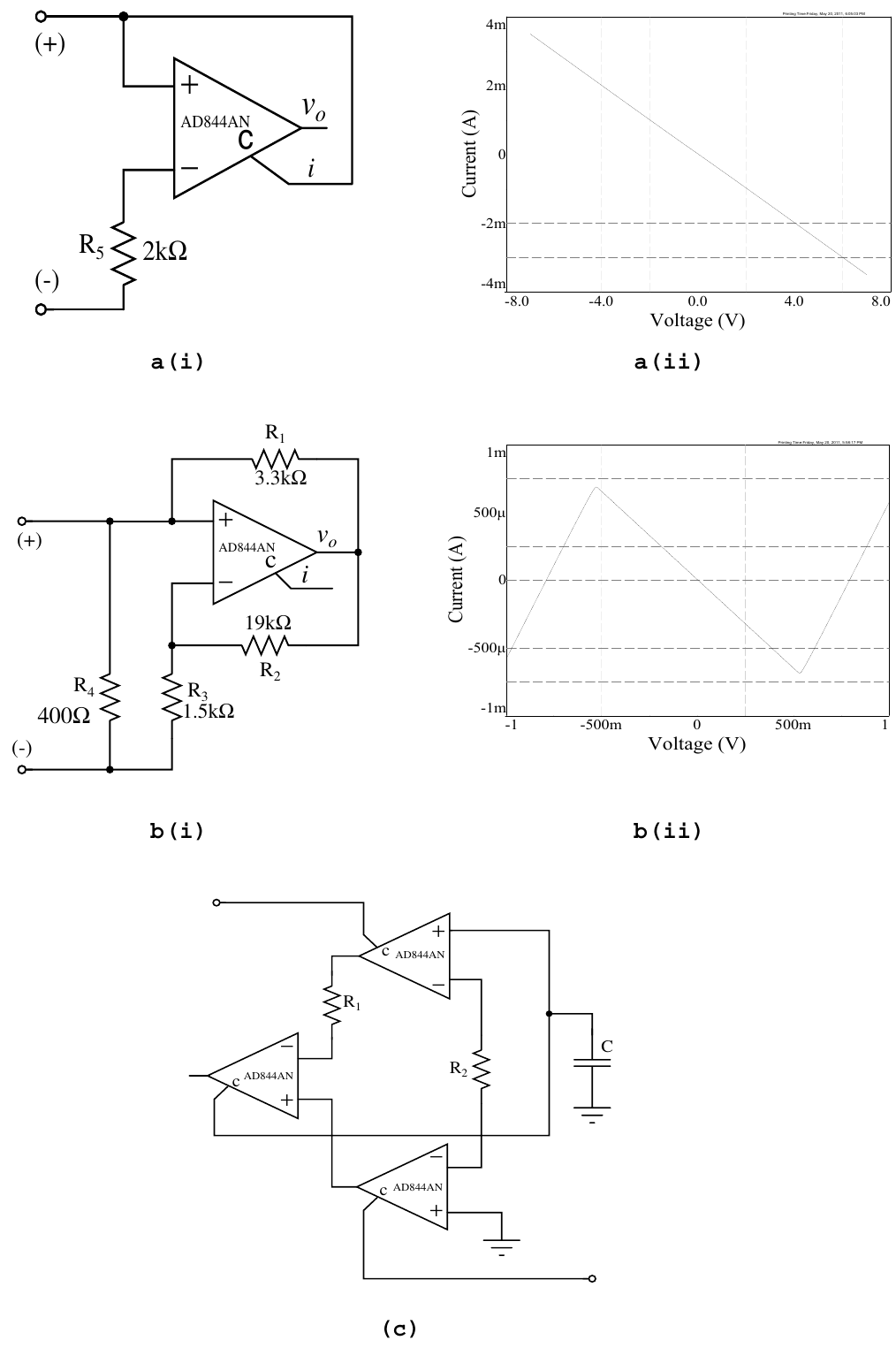,width=4.5in}}
\vspace*{6pt}
\caption{(i) Current feedback operational amplifier (CFOA) based circuit realization and (ii) the simulated ($v-i$) characteristics of (a) Negative conductor ($g_{N}$), (b) Nonlinear resistor ($R_{N}$) and (c) the synthetic floating inductance simulator ($L_{eq}$) realized using CFOAs.}
\end{figure}

\newpage
\begin{figure}[th]
\centerline{\psfig{file=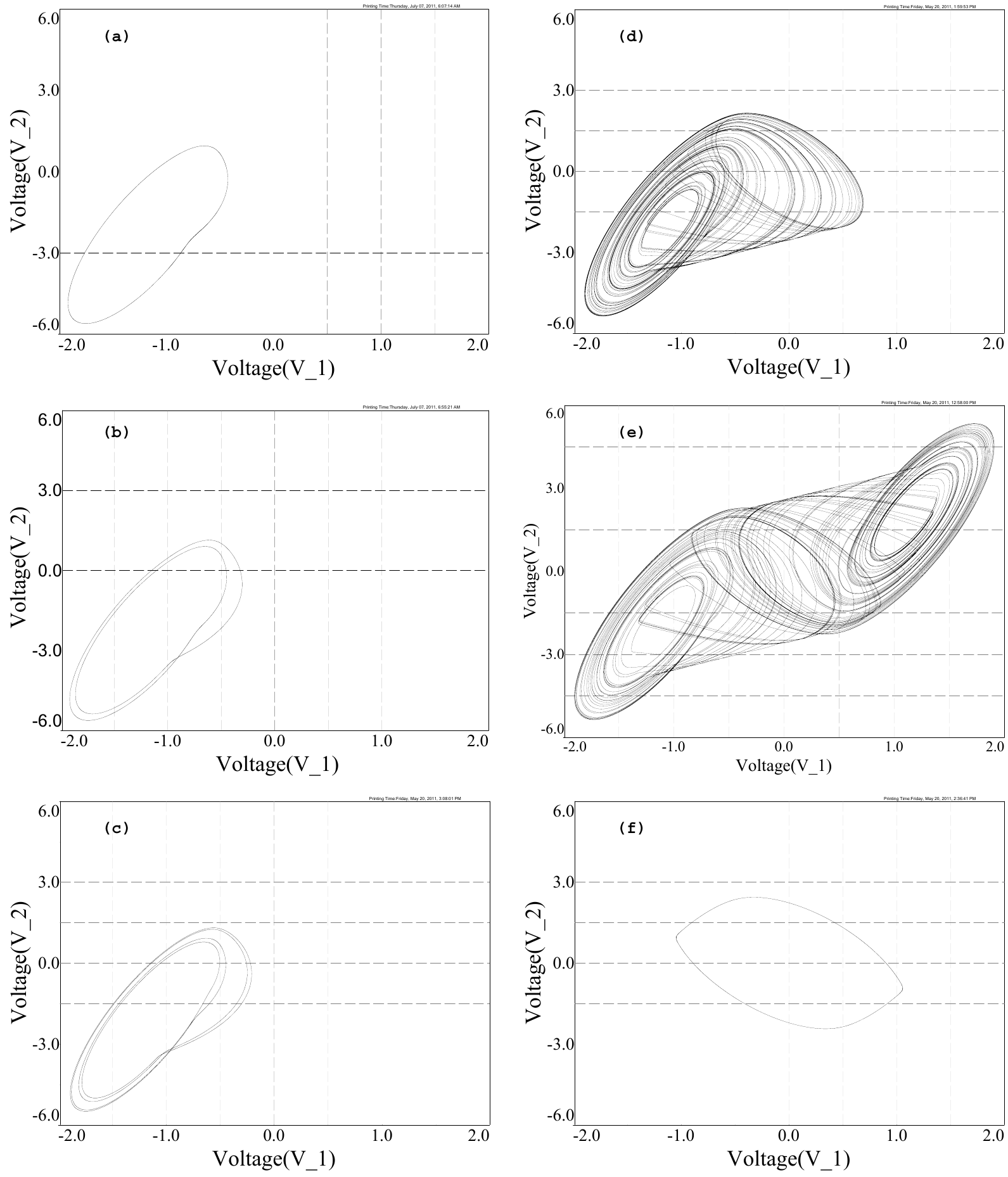,width=5.0in}}
\vspace*{6pt}
\caption{The period doubling scenario in the improved canonical Chua's circuit shown in Fig. (4). The phase portraits in the ($v_{1}-v_{2}$) plane, for the fixed values of the circuit elements chosen as $C_{1}=7.5$ pF, $C_{2}=60$ pF and $L_{eq}=180~\mu$H, by decreasing resistance ($R$) values: (a) period - 1 limit cycle ($R=985~\Omega$), (b) period - 2 limit cycle ($R=968~\Omega$), (c) period - 4 limit cycle ($R=960~\Omega$), (d) single band chaotic attractor ($R=900~\Omega$), (e) double band chaotic attractor ($R=850~\Omega$) and (f) boundary ($R=805~\Omega$).}
\label{fig_phase}
\end{figure}

\newpage
\begin{figure}[th]
\centerline{\psfig{file=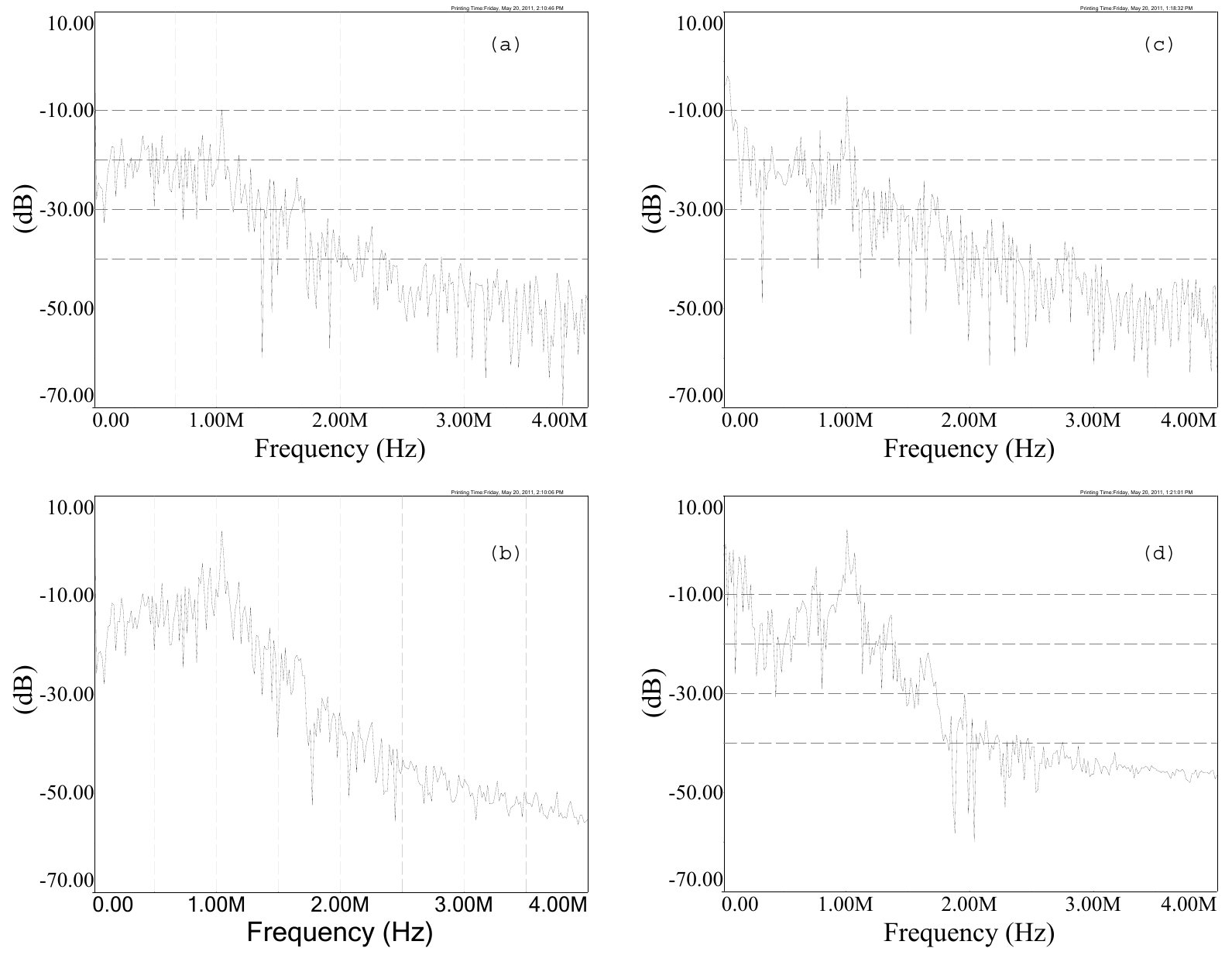,width=6.0in}}
\vspace*{6pt}
\caption{Power spectrum for the time series of the variables $v_{1}$ and $v_{2}$ of the canonical Chua's circuit. The dominant frequency is centered around (a-b) 1.0313 MHz for the single band chaos and (c-d) 1.000 MHz for the double band chaos.}
\end{figure}

\newpage
\begin{figure}[th]
\centerline{\psfig{file=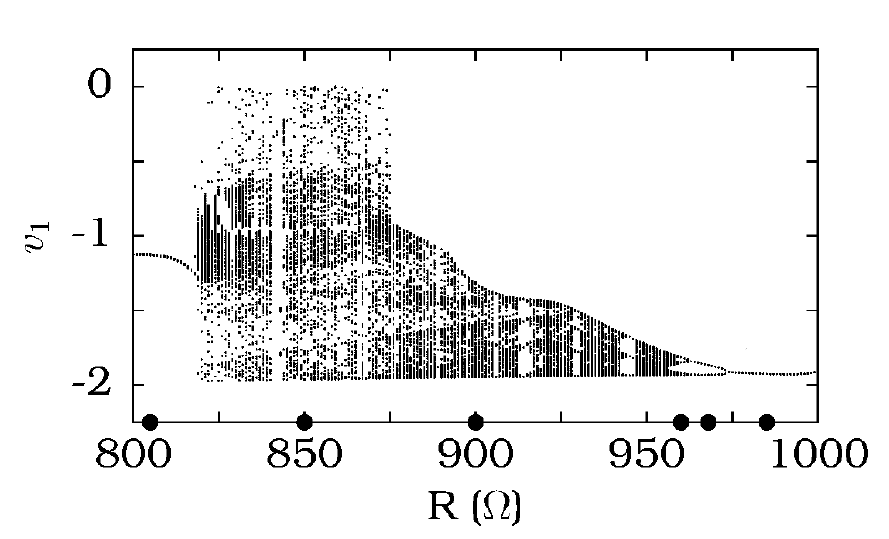,width=4.0in}}
\vspace*{6pt}
\caption{The one parameter bifurcation diagram generated in the ($v_1-R$) plane for the time series of the variable $v_{1}$. The filled circles in the horizontal axis denotes the parameters used to plot the phase portraits in Fig. \ref{fig_phase}.}
\label{fig_bif}
\end{figure}

\newpage
\begin{figure}[th]
\centerline{\psfig{file=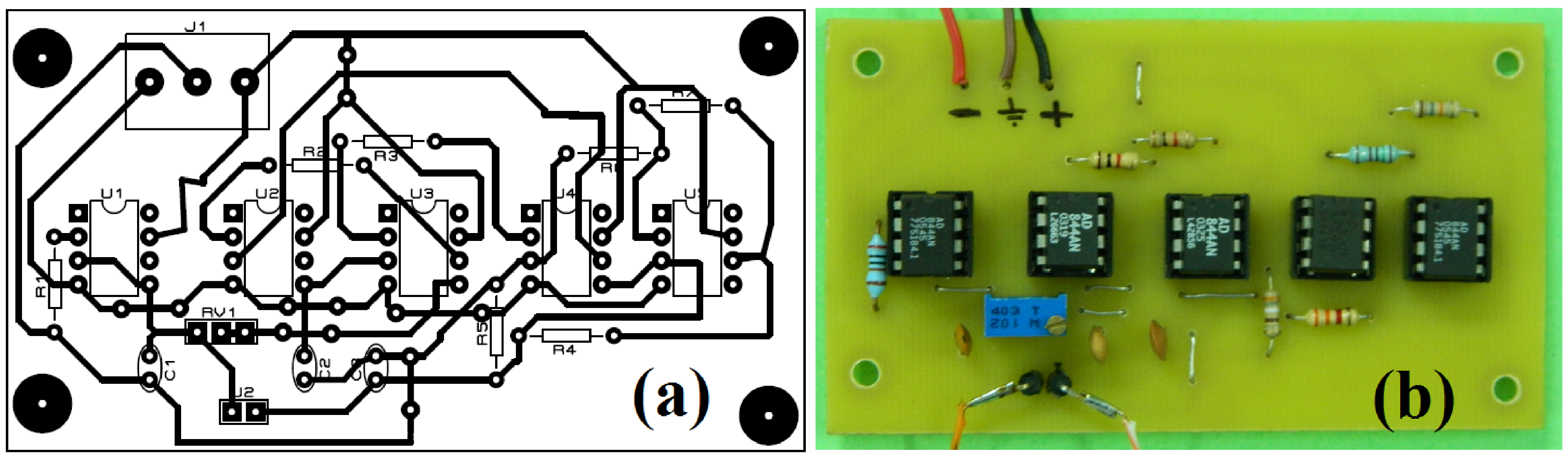,width=\columnwidth}}
\vspace*{6pt}
\caption{(color online) The canonical Chua's circuit (a) PCB layout and (b) the PCB circuit itself.}
\label{fig9}
\end{figure}

\newpage
\begin{figure}[th]
\centerline{\psfig{file=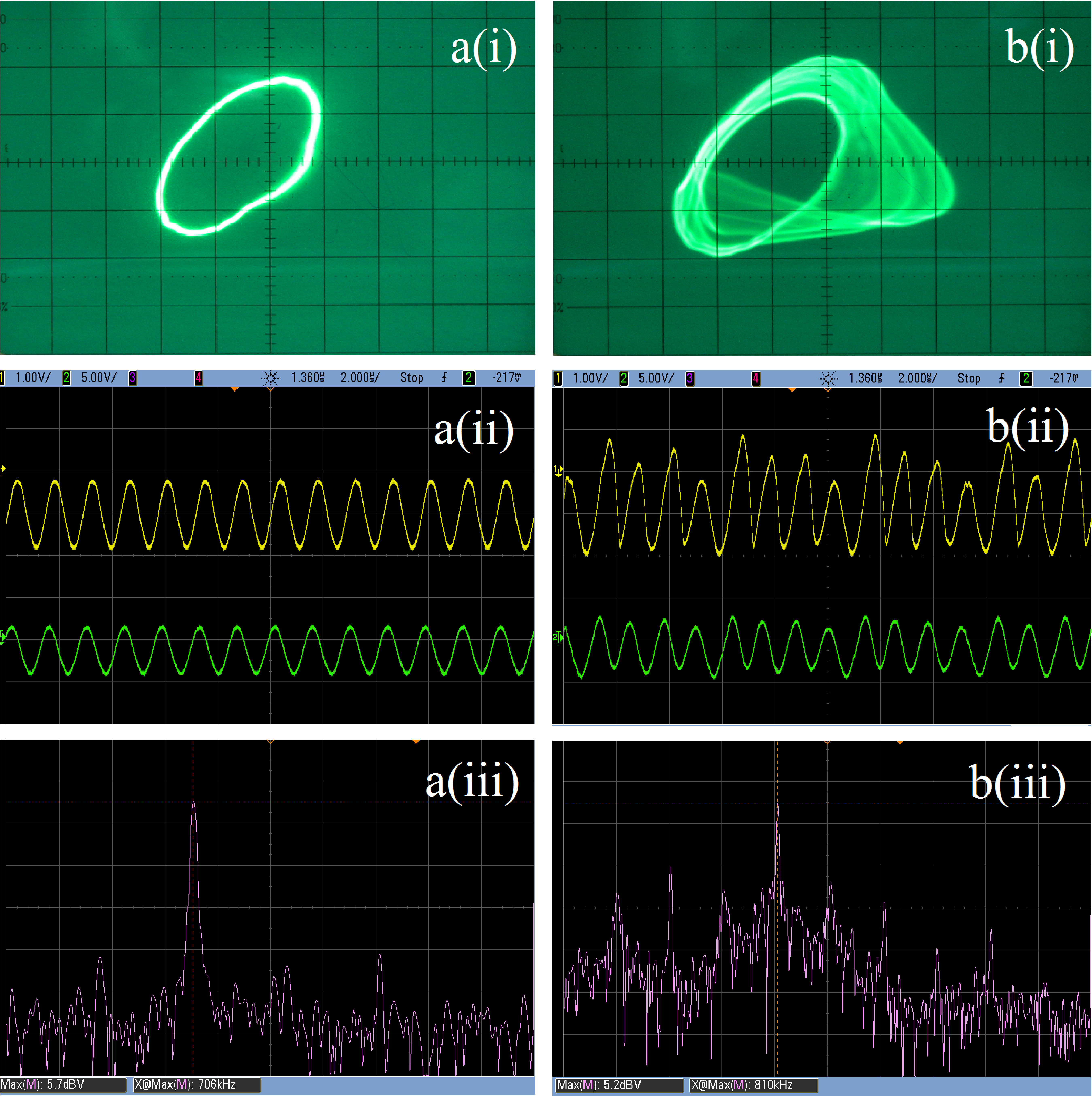,width=\columnwidth}}
\vspace*{6pt}
\caption{(color online) The phase portrait in ($v_1-v_2$) plane for $a(i)$ periodic $b(i)$ chaotic motion for the value of the resistor $R$ as $471~\Omega$ and $363~\Omega$ respectively. The time series for the variable $v_1$ (yellow) and $v_2$ (green) are shown in $a(ii)$ for periodic and $b(ii)$ d for chaotic motion. Scale:- $x-$axis: $2\mu$S/Div; $y-$axis: (a) voltage $v_1$-$1$V/Div and (b) voltage $v_2$-$5$V/Div. The power spectrum of $v_1$ showing the dominant peak at 706kHz for periodic and 810 kHz for chaotic motion are depicted in $a(iii)$ and $b(iii)$ respectively. The amplitude of these two dominant peaks are noted as 5.7 dBV and 5.2 dBV.}
\label{fig10}
\end{figure}

\end{document}